\documentstyle[12pt]{article}

\def\half{{\textstyle{1\over2}}}

\def\fourth{{\textstyle{1\over4}}}
\def\>{\rangle}
\def\<{\langle}

\def\doeack{This work  was supported in part by the Department of Energy,
Nuclear
      Physics Division, under
      contracts W-31-109-ENG-38.}
\newcommand{\sect}[1]{\addtocounter{section}{1}\begin{center}
        \normalsize{\bf\arabic{section}.~ #1}\end{center}\vspace*{4mm}}
\newcommand{\sectA}[1]{\addtocounter{section}{1}\begin{center}
        \normalsize{\bf #1}\end{center}\vspace*{4mm}}

\def\beq{\begin{equation}}
\def\eeq{\end{equation}}
\def\beqarray{\begin{eqnarray}}
\def\eeqarray{\end{eqnarray}}

\def\V{{\cal V}}
\def\U{{\cal U}}

\date{  26.08.93}
\begin{document}
\title{Electromagnetic Currents  and the Blankenbecler-Sugar Equation}
\author{F. Coester$^1$ and D.O. Riska$^2$}
\maketitle
\centerline{$^1$ \it Physics Division, Argonne National Laboratory,
Argonne, IL 60439-4843, USA}

\centerline{$^2$ \it Department of Physics, University of Helsinki,
00170 Helsinki, Finland}
\setcounter{page}{0}
\vspace{1cm}

\centerline{\bf Abstract}

The effective electromagnetic current density  for a
two-nucleon system that is described by the Blankenbecler-Sugar
equation is derived.
In addition to the single nucleon currents there are exchange currents of
two different  origins.
The first is the exchange current that is required to compensate for
the violation of the continuity equation in the impulse approximation.
The second is an exchange current, which arises in
the quasipotential
reduction from the Bethe-Salpeter equation, and which represents
effects of suppressed  degrees of freedom. Explicit general
expressions are given for both of these exchange currents. The results are
applicable to both elastic and inelastic processes.
\newpage
\small
\sect{Introduction}
\normalsize
\vspace{0.3cm}

The Blankenbecler-Sugar (BSLT) equation \cite{Sugar,Logun} provides a
convenient
three-dimensional quasipotential framework for describing two-body
bound states by covariant amplitudes defined as functionals
of local quantum fields. The relation between the underlying Bethe-Salpeter
(BS)
wave function and the invariant electromagnetic  form factors \cite{Mandel}
 determines
the effective current kernel, which yields the
electromagnetic form factors when
folded with the covariant deuteron vertex. The relation of the form
factors to the BSLT wave functions rests on the observation that the deuteron
vertex obtained by applying the quasipotential to the BSLT wave function is
 identical
to the vertex obtained by applying the BS potential to the BS wave
 function \cite{Jaus}.
In the impulse approximation this observation is  sufficient  for the
 calculation of form
factors in terms BSLT wave functions. The elimination of the explicit
negative-energy components of the Dirac
spinors by the quasipotential reduction results in the appearance of effective
two-nucleon exchange currents. It
is the aim of this paper to derive explicit expressions for these
exchange currents.

Lorentz invariant equal-time constraints and the well-known
partial fraction decomposition of the single nucleon propagator  facilitate a
detailed  comparison with conventional exchange current
phenomenology, which  is based on a reduction of Hamiltonian dynamics of
nucleons,
antinucleons and mesons to  effective Hamiltonians and current operators
acting on the two-nucleon  Hilbert space \cite{Friar}. In this context
anti-nucleon degrees of freedom  can be described by including negative-energy
Dirac
spinors among the nucleon degrees of freedom. The effective exchange currents
resulting from the elimination of these negative energy-Dirac spinors can be
compared to the effective exchange currents arising from the elimination of
negative-energy Dirac spinors in quasipotential reductions.

In Section  2 we summarize the main features of the exact
Mandelstam-Bethe-Salpeter relations and associated exact quasipotential
reductions.
These features yield exact expressions for
current matrix elements based on axiomatic properties of local quantum field
operators.
The three- and four-point vertex functions obtain equivalently as solutions of
BS or BSLT equations.
In practice an assumed covariant quasipotential and the three-point
current vertex form the  input for the BSLT phenomenology. The relation between
the Bethe-Salpeter phenomenology and quantum field theory is merely formal.
Questions of consistency of common practices with basic principles are outside
the scope of this paper.
In order to make contact with conventional exchange current phenomenology,
which is based on nonrelativistic quantum mechanics,
we decompose the  BS impulse
approximation for the
current matrix element  into a reduced   BSLT impulse approximation and an
effective exchange current.

The impulse approximation for the two-nucleon current vertex does not, in
general, satisfy current conservation even at the level of the $BS$
equation. This violation  of current conservation in the  impulse approximation
must be compensated by additional  two-nucleon currents.
In the framework of the Bethe-Salpeter formalism
Gross and Riska  \cite{Gross} have shown  how to exploit the Ward-Takahashi
identities
\cite{Ward,Taka} of the three-point  current vertex  to restore current
 conservation
by a  minimal two-nucleon  current constructed as a functional of the BS
potential. In Section 3  we derive a  corresponding minimal two-body  current
 kernel as a functional of the BSLT potential.

In Section 4 we describe the relation
between the exchange currents for the BSLT framework and the
conventional phenomenological exchange current operators and show
that in the static limit there are terms that agree with
the pair current operators of the
quantum mechanical formulation.
Section 5 contains a concluding
discussion.\\


\sect{The effective current kernel}
\vspace{0.3cm}
\noindent{\bf 2.1 The Mandelstam-Bethe-Salpeter relations}
\vspace{0.2cm}

Matrix elements $\<P_f|J^\mu(x)|P_i\>$
of the electromagnetic current-density
operator $J^\mu(x)$ between the eigenstates $|P_i\>$ and $|P_f\>$ of
the total momentum operator are the quantities directly related to
cross sections
for elastic and inelastic electron scattering as well as  photo-disintegration.
In the Bethe-Salpeter framework these matrix elements
can be expressed in terms of Bethe-Salpeter wave functions $\Psi_P
(p_1, p_2)$ and a current kernel $K_\mu(p_1',p_2';p_1,p_2)$ as
\cite{Mandel,Coester}
\newpage
\beqarray
\<P_f|J^\mu(0)|P_i\>&=&\int d^4 p_1^{'} d^4 p_2^{'} d^4 p_1
d^4p_2\cr\cr
&&\times \bar \Psi_{P_f} (p_1^{'}, p_2^{'}) K^\mu(p'_1, p'_2;p_1, p_2)
\Psi_{P_i} (p_1, p_2)\;.\label{BSC}
\eeqarray
Here the Bethe-Salpeter wave functions $\Psi_{P} (p_1, p_2)$  are
Fourier transforms
of matrix elements of time ordered products of nucleon field
operators of the form
$\<0 \vert T\{\psi(x_1)\psi(x_2)\}\vert P\>$.
These can be extracted from the four-point Green function
$G(x_1',x_2';x_2,x_1)$, which is
defined as the vacuum expectation value
of a time ordered product of local operator valued distributions as
\beq
G(x_1',x_2';x_2,x_1):=
\<0 |T\{\psi(x_1')\psi(x_2')\bar \psi(x_2)\bar \psi(x_1)\}|0\>\; .
\label{BSG}
\eeq
The Green function with  the assumed spectrum of the  four-momentum operator
determines the Bethe-Salpeter wave functions of bound states.
Scattering states are obtained by the large time limits of the field operators
\cite{Schweber}.

The current kernel $K^\mu(p_1',p_2';p_1,p_2)$ is   related to the
Fourier transform of the five-point current vertex,
\beq
R^\mu_5(x_1',x_2';x_1,x_2):=
\<0 |T\{\psi_1(x_1')\psi_2(x_2')J^\mu(0)\bar \psi_1(x_2)\bar \psi(x_1)\}|0\>
\; ,
\label{BSR}
\eeq
by the definition
\beq
R^\mu_5 = G K^\mu G\; .
\label{GKG}
\eeq
Here the Green function  is treated as a formal integral
operator, for which an inverse is a assumed to exist.

The Bethe-Salpeter equation for a bound state wave function $\Psi$ may
be written formally as
\beq
G^{-1}\Psi=( G_f^{-1} -  {\cal U} )\Psi=0 \; ,
\label{BSE}
\eeq
where $G_f$ is the Green function of two isolated nucleons:
\beq
G_f(x_1',x_2';x_2,x_1):=\<0 |T\{\psi_1(x_1')\bar \psi_1(x_1)\}|0\>
\<0 |T\{\psi_2(x_2')\bar \psi_2(x_2)\}|0\>\; ,
\eeq
and the Bethe-Salpeter potential ${\cal U}$ is defined  as the difference
between the full and free inverse Green functions:
\beq
{\cal U}:=  G_f^{-1}-G^{-1}\; .
\label{UU}
\eeq

Functions defined as vacuum expectation values of products  of local
field operators become products of vacuum expectation values for large
spatial separation of the points. The ``truncated vacuum expectation
values'' are  defined as the parts which vanish  for any separation
of the points into widely separated clusters. Thus
\beq
G(x_1',x_2';x_2,x_1)=G_f(x_1',x_2';x_2,x_1)
+\<0 |T\{\psi_1(x_1')\psi_2(x_2')\bar \psi_2(x_2)\bar \psi_1(x_1)\}|0\>_T\; ,
\eeq
where
 the subscript $T$ indicates the truncated part and the two particle are
assumed to be distinguishable.
For the five-point current vertex  a similar cluster
decomposition involves products of the three-point current vertex,
\beq
R^\mu_3(x_1';x_1):=\<0 |T\{\psi(x_1')J^\mu(0)\bar \psi(x_1)\}|0\>,
\eeq
with the single-nucleon Green function,
\beq
G_1(x_2';x_2):=\<0 |T\{\psi(x_2')\bar \psi(x_2)\}|0\>\; ,
\eeq
and a truncated remainder, which vanishes in the absence of interactions:
\beqarray
&&R^\mu_5(x_1',x_2';x_1,x_2)=R^\mu_3(x_1';x_1)G_1(x_2';x_2)+
R^\mu_3(x_2';x_2)G_1(x_1';x_1)\cr\cr
&&+\<0 |T\{\psi_1(x_1')\psi_2(x_2')J^\mu(0)\bar \psi_2(x_2)\bar
 \psi_2(x_1)\}|0\>_T
\; .
\label{BSRC}
\eeqarray
In the ``relativistic impulse approximation''
the current vertex $R^\mu_5$ is approximated by the first two terms on the
right
hand side of eq. (\ref{BSRC}),
\beq
R^{IA}_{5\mu}= R_{3,\mu}(1)G_1(2)+ R_{3,\mu}(2)G_1(1)
= G_1(1)K^{(1)}_\mu G_f+ G_1(2)K^{(2)}_\mu G_f\; .
\label{RIA}
\eeq
The remainder $R_\mu^{EX}:=R_\mu-R_\mu^{IA}$ represents
a two-nucleon ``exchange'' (or ``interaction'') current,
which vanishes in the absence of interactions.

As a consequence of translational invariance
the Fourier transforms  of the two-nucleon Green functions $G$ and $G_f$
are proportional to delta functions as follows
\beq
G(p_1',p_2';p_1,p_2)= \delta^{(4)}(P'-P)G_P(p';p),
\eeq
\beq
G_f(p_1',p_2';p_1,p_2)=\delta^{(4)}(P'-P)\delta^{(4)}(p'-p)G_0(p,P)\; .
\eeq
Here the total and relative momenta are denoted by $P$ and $p$ respectively:
\beq
P:= p_1+p_2\; ,\qquad p:={1\over 2}(p_1-p_2)\; .
\eeq

In a quantum field theory with non-trivial interactions the two-point
Green function $G_1(x',x)$ cannot be equal to the free-field Green function.
In other words the continuous part of the Lehmann weight cannot vanish.
It is common practice in nuclear BS phenomenology to ignore the
continuous part of the Lehmann weight. We will adhere to this practice
in the following.
The free-nucleon propagator $G_0(p,P)$ is then defined as the product,
$G_0(p,P):=S(p_1)S(p_2)$, of the single-nucleon
propagators, $S(p_k)$, $(k=1,2)$:
\beq
S(p_k):={-1\over 2\pi i[\gamma^{(k)}\cdot p_k +(m -i\epsilon/2m)]}=
{\gamma^{(k)}\cdot p_k-m\over2\pi i( p_k^2+m^2-i\epsilon)}\; .
\label{PROP}
\eeq

{}From the definitions (\ref{UU}) of the BS potential and
\beq
T_P(p';p):= G_0(p',P)^{-1}G_P(p';p)G_0(p,P)- \delta^4(p'-p)G_0(p,P)^{-1}
\eeq
of the four-point vertex $T_P(p';p)$ follows the
Bethe-Salpeter equation
\beq
T_P(p';p)= \U_P(p';p) + \int d^4 p''\U_P(p';p'')G_0(p'',P) T_P(p'';p)\; .
\label{BSET}
\eeq
In practice an assumed Bethe-Salpeter
potential $\U_P$ determines the four-point vertex $T_P(p';p)$ and hence the
Green function
\beq
G_P(p';p)= \delta^4(p'-p)G_0(p,P)+G_0(p',P)T_P(p';p)G_0(p,P)\; .
\label{GT}
\eeq

It follows from the relation of the Green function to products of local field
operators that the BS wave function of the deuteron,
\beq
\Psi_{P_i}(p_1,p_2)=\delta^{(4)}(P_i-P)\chi_P(p)\; ,
\eeq
is related to a pole of the Green function $G_P$ at the deuteron mass:
\beq
\lim_{P^2\to -M_d^2} (P^2+M_d^2)T_P(p',p)= {1\over 2\pi i}\Gamma_P(p')
\bar \Gamma_P(p)\; ,
\label{POLE}
\eeq
where $\Gamma_P(p:= G_0^{-1}(p,P)\chi_P(p)$.
As a consequence the deuteron vertex $\Gamma_P(p)$ satisfies the BS equation
\beq
\Gamma_P(p)=\int d^4p' \U_P(p;p')G_0(p',P)\Gamma_P(p')
\label{BSEG}
\eeq
and the ``normalization condition''\cite{Lurie}:
\beq
-{1\over 2\pi i}\int d^4p'\int d^4 p \bar \Gamma_P(p')G_0(p',P)
{\partial G_P^{-1}(p',p)\over \partial P^0}G_0(p,P)\Gamma_P(p) =2P^0\; .
\label{NN}
\eeq
Here the covariant normalization convention for momentum eigenstates
\beq
\int d^4P'\, \vert P'\> \delta(P{'}^2+M^2) \< P'\vert P\> = \vert P\>,
\eeq
has been used.
The relations (\ref{POLE}) and (\ref{NN}) make it possible to extract
the matrix element
$\<P_f|J^\mu(0)|P_i\>$  for elastic  electron deuteron scattering
from eq. (\ref{GKG}) \cite{Mandel,Coester}. The result is $$the Mandelstam
relation
\beq
\<P_f|J^\mu(0)|P_i\>=\int d^4 p' \int d^4 p \bar\Gamma_{P_f}(p')
G_0(p',P_f) K^\mu(P_f,p';P_i,p)G_0(p,P_i)\Gamma_{P_i}(p)\; .
\label{JGAM}
\eeq

For inelastic processes the construction of current matrix elements is based
on the observation that the field operators $\psi_1(x_1)$  $\psi_2(x_2)$
converge
to the ``out'' fields \cite{Schweber} for large positive times.
The action of a product of ``out'' fields on the vacuum generates scattering
states $\vert \vec P_f ,\vec p_f\>$.
Current matrix elements  for inelastic processes can be obtained by taking the
large-time limit,
\beqarray
&&\<\vec p_f,\vec P_f\vert J^\mu(0)|P_i\>=
\int dp_{1f}^0\int dp_{2f}^0
\lim_{t_1\to \infty}\lim_{t_2\to \infty} e^{i(E_{1f}-p^0_{1f})t_1}
e^{i(E_{2f}-p^0_{2f})t_2}\cr\cr
&&\times\int d^4 p'\int d^4 p G_{P_f}(p_f;p')K^\mu(p',P_f;p, P_i)G_0(p,P_i)
\Gamma_{P_i}(p)\; ,
\label{OUT}
\eeqarray
where $E_{kf}:= {\sqrt{\vec p_{kf}^2+m^2}}$.
The large time limit of the free Green function can be taken explicitly,
\beqarray
&&\lim_{t_1\to \infty}\lim_{t_2\to \infty} e^{i(E_1-p^0_1)t_1}
e^{i(E_2-p^0_2)t_2}G_0(p_1,p_2)\cr\cr
&&=(\gamma^{(1)}\cdot p_1-m)(\gamma^{(2)}\cdot p_2-m)\theta(p_1^0)\theta(p_2^0)
\delta(p_1^2+m^2)\delta(p_2^2+m^2)\; .\cr\cr
&&
\label{LLIM}
\eeqarray
Thus it follows  from eqs. (\ref{OUT}) and (\ref{GT}) that
\newpage
\beqarray
&&\<\vec p_f,\vec P_f\vert J^\mu(0)|P_i\>=
{(\gamma^{(1)}\cdot p_{1f}-m)(\gamma^{(2)}\cdot p_{2f}-m)\over 4E_{1f}E_{2f}}
\cr\cr &&
\Biggl\{\int d^4 p K^\mu(p_f,P_f;p,P_i) G_0(p,P_i)\Gamma_{P_i}(p)\cr\cr
&&+\int d^4p'\int d^4p T_{P_f}(p_f;p')
G_0(p',P_f) K^\mu(P_f,p';P_i,p)G_0(p,P_i)\Gamma_{P_i}(p)\Biggr\}.\cr\cr
&&
\eeqarray
In the impulse approximation $K \to K^{IA}$ we have, with $Q:=P_f-P_i$,
\beqarray
&&G_0(p',P_f)K^{IA}(P_f,p';P_i,p)G_0(p,P_i)=\cr\cr
&&\delta(p'-p-\half
Q)S(p'_1)K^{(1)}(p_1';p_1)S(p_1) S(p_2)+(1 \leftrightarrow 2) \; .
\label{KIA}
\eeqarray
The term $(1 \leftrightarrow 2)$  is equal
to the first term with
all the coordinates of the two nucleons interchanged.\\

For elastic scattering the current matrix elements in the impulse
approximation are therefore
\beqarray
&&\<P_f|J^{IA}_\mu(0)|P_i\>= \int d^4 p'\int d^4 p \delta^{(4)}(p'-p-\half
Q)\cr\cr
&&\times \bar\Gamma_{P_f}(p')
 S(p_1')K^{(1)}_\mu(p_1';p_1)G_0(p,P_i)\Gamma_{P_i}(p)+(1 \leftrightarrow 2) \;
,
\label{JIA}
\eeqarray
whereas for inelastic scattering we have
\beqarray
&&\<P_f, \tilde p_f|J^{IA}_\mu(0)|P_i\>={(\gamma^{(1)}\cdot
p_{1f}-m)(\gamma^{(2)}
\cdot p_{2f}-m)\over 4\omega_f^2}\cr\cr
&&\times \Biggl\{\int d^4 p \delta(p_f- p-\half Q)K^{(1)}_\mu
(p_{1f};p_1) S(p_1)
\Gamma_{P_i}(p)\cr\cr
&&+\int d^4p \int d^4 p'\delta(p'- p-\half Q) \,T_{P_f}(p_f;p')S(p_1')
K^{(1)}_\mu(p_1';p_1)G_0(p,P_i)
\Gamma_{P_i}(p)\Biggr\}\; .\cr\cr
&&
\label{JIAC}
\eeqarray
The second term has
the same formal structure as the elastic matrix element.

When $G_P^{-1}$  is replaced in the normalization condition (\ref{NN})
by the corresponding free-field function
$S^{-1}(p_1)S^{-1}(p_2)$  that normalization condition
reduces to
\beq
\half \int d^4 p \bar \Gamma_P(p)\left[S(p_1')\gamma^{(1)0}G_0(p,P)
+(1 \leftrightarrow 2)\right]\Gamma_P(p) =2 P^0 \; .
\label{NBS}
\eeq
In this  form  the normalization of the wave function
manifestly guarantees the correct value of the total charge in the impulse
approximation.\\

For
${p_1'}^2=p_1^2=-m^2$ the single nucleon
current vertex $K^{(1)}_\mu(p_1';p_1)$ can be expressed in terms of isospin
dependent Dirac
and Pauli form factors as
\beq
K^{(1)\mu}= \tilde{F}^{(1)}_1(Q^2)\gamma^{(1)\mu}
+{1\over 4m}[\gamma^{(1)\mu},\gamma^{(1)}\cdot Q]\tilde{F}
^{(1)}_2(Q^2)\; .
\label{FAC}
\eeq
Here $\tilde{F}_1:=\half F_1^S+\half \tau^3 F_1^V$ and
$\tilde{F}_2:=\half F_2^V+\half \tau^3 F_2^V$.
The usual practice of assuming the (\ref{FAC}) for all values of
$p_1$ and  $p_1'$ violates the Ward-Takahashi identity \cite{Ward,Taka},
which is required by the gauge covariance of the single nucleon propagator.
Even after this  defect is removed the impulse approximation (\ref{JIA})
violates current conservation for typical nuclear interactions \cite{Gross}.
The exchange currents that are required
to cure this deficiency will be discussed in Sec. 3.

For the evaluation of the integrals (\ref{JIA}) and (\ref{JIAC})
it is convenient to use as an integration variable the Lorentz invariant
relative
energy, $k^0$ of the target,
\beq
k^0:=- p\cdot \hat P_i\;,\qquad  \hat P_i:=P_i/M_i\; ,
\eeq
and the four-vector $\tilde p$,
\beq
\tilde p :=p-k^0\hat P_i,
\eeq
which is  the transverse component of the relative momentum.
The corresponding variables of the final
state,
\beq
{k'}^0:=- p'\cdot \hat P_f\;,\qquad  \hat P_f:=P_f/M_f\; ,\qquad
\tilde p' :=p'-{k'}^0\hat P_f,
\eeq
are related by the  momentum constraint $p'=p+\half Q$.
For elastic scattering, $M:=M_i=M_f$, we have the relations
\newpage
\beqarray
&&{k'}^0-k^0=-{\tilde p'\cdot Q +\tilde p\cdot Q\over 2M(1+\eta)},\cr\cr
&&\tilde p'- \tilde p= \half Q\left(1-{{k'}^0+k^0\over M}\right)
-{{k'}^0- k^0\over M}\, \half (P_f+P_i),
\label{CON}
\eeqarray
and
\beqarray
&&\tilde p'\cdot Q -\tilde p\cdot Q =\half Q^2\left(1-{{k'}^0+k^0\over
M}\right)
 \;,\cr\cr
&&\tilde {p'}^2-\tilde p^2 =({k'}^0-k^0)(k^0+{k'}^0-M)\; ,
\label{DOM}
\eeqarray
where we use the notation $\eta :=Q^2/4M^2$.
For inelastic processes, for which
$M_f\not=M_i$, the relations (\ref{CON}) and (\ref{DOM}), which give
${k'}^0$ and $\tilde p'\cdot Q$ as functions of $k^0$ and $\tilde p\cdot Q$,
 must be  replaced by
\beq
{k'}^0M_f-k^0M_i = \left({k'}^0M_f+k^0M_i\right){M_f^2-M_i^2\over M_f^2+M_i^2}
-{2M_f^2M_i^2(\tilde p\cdot Q +\tilde p'\cdot Q)
\over (M_f^2+M_i^2)(M_f^2+M_i^2+\half Q^2)}\; ,
\eeq
and
\beq
\tilde p'\cdot Q -\tilde p \cdot Q = \half Q^2\left(1-{{k'}^0\over M_f}
-{k^0\over M_i}\right)+ \half(M_f^2-M_i^2)
\left(1-{{k'}^0\over M_f}+{k^0\over M_i}\right)\; .
\eeq

The four-vector $\tilde p$ is
determined by three independent variables, the choice of which is not important
for the following development. For completeness we nevertheless list some
appropriate choices and their relationships in the Appendix.
Any Lorentz invariant function
of the two four-momenta $p$ and $P$ is a function of $k^0$, $M$ and
$\omega:={\sqrt{\tilde p^2 +m^2}}$.

The Blankenbecler-Sugar constraint restricts $k^0$ to $0$ while the invariant
equal-time constraint, $P\cdot (x_1-x_2)=0$ is realized by integration over
$k^0$. These features play a crucial role in the interpretations of the
impulse approximation for the current
matrix elements.
\newpage
\noindent{\bf 2.2 Quasipotential reductions}
\vspace{0.2cm}

Quasipotential reductions of the Bethe-Salpeter formalism are useful for
the explicit construction of
deuteron vertex $\Gamma_P(p)$ and the  four-point vertex
$T_P(p',p)$. These reductions are generated by
a two-nucleon propagator $g_0(p,P)$ that constrains the relative momentum
to a three-dimensional manifold. We will restrict ourselves to
quasipotential
propagators that involve the BSLT constraint $p\cdot P =0$,
\beq
g_0(p,P)=\delta(k^0)\hat g_0(\tilde p,P)\; .
\eeq

The Bethe-Salpeter  equations (\ref{BSET}) for the  four-point vertex,
and (\ref{BSEG}) for the  deuteron vertex,
are equivalent to the quasipotential equations
\beq
T_P(p';p)= \V_P(p';p) + \int d^4 p''\V_P(p';p'')g_0(p'',P) T_P(p'';p)\; .
\label{TTV}
\eeq
and
\beq
\Gamma_P(p)= \int d^4 p'{\cal V}_P(p;p') g_0(p',P) \Gamma_P(p').
\label{VV}
\eeq
Here the quasipotential ${\cal V}_P$ is determined by the BS potential
$\U_P$  by the integral equation
\beq
 \V_P(p';p)={\cal U}_P(p';p)+\int d^4p''{\cal V}_P(p';p'')
[G_0(p'',P) - g_0(p'',P)]\U_P(p'';p).
\label{VU}
\eeq
{}From the BS equation (\ref{BSEG}) and eq.(\ref{VU}) follows the identity
\beq
\int d^4 p'{\cal U}_P(p;p') G_0(p',P) \Gamma_P(p')
\equiv\int d^4 p'{\cal V}_P(p;p') g_0(p',P) \Gamma_P(p')\; .
\label{EQG}
\eeq
It should be emphasized that the vertices  $\Gamma_P(p)$  and
$T_P(p';p)$ that are obtained by solving
equations (\ref{TTV}) or (\ref{BSET}), and  (\ref{VV}) or (\ref{BSEG})
are identical when the potentials are related
by eq.~(\ref{VU}). Either potential may be considered the primary quantity
constructed from appropriate Feynman diagrams. The other is then defined
by eq.~(\ref{VU}).

The quasi-potential reduction   provides
not only a convenient device for the construction of the vertices,
but it is also designed to provide contact with the nonrelativistic quantum
mechanical description and
conventional exchange-current phenomenology. For that reason we wish to express
the matrix element $\<P_f\vert J_\mu^{IA}(0)\vert P_i\>$
as a matrix element with respect to  BSLT wave
functions.
The initial and final BSLT deuteron wave functions,
\beqarray
&&\varphi_{P_i}(p):= \hat g_0(\tilde p){\Gamma_{P_i}(p)\over {\sqrt{2
 P_i^0}}}\; ,\cr\cr\cr
&&\bar \varphi_{P_f}( p')=-{\bar\Gamma_{P_f}( p')
\over {\sqrt{2 P_f^0}}} \hat g_0(\tilde p')\; ,
\label{WFNBS}
\eeqarray
and the  scattered wave
\beq
T_{P_f}(p_f; p')\hat g_0(\tilde p')\; ,
\eeq
are defined for all values of $p$ and $p'$.
This is important because
the initial and final BSLT constraints $k^0=0$ and
${k'}^0 =0$ are incompatible \cite{Devine} with the momentum constraint
$p'-p-\half Q=0$,
which implies the relations (\ref{CON}).
Thus  the current matrix element will necessarily
involve values of the BSLT wave functions for nonvanishing values of
$k^0$.

When the quasipotential is an ``instantaneous'' interaction, i.e. when
$\V_P(p';p) = \V(\tilde p'; \tilde p)$, it is evident from eq. (\ref{VV})
that the  deuteron vertex
and the BSLT wave function are independent of $k^0$.
In general the  vertices $\Gamma_P$ and $T_P$ are
slowly varying functions of $k^0$, which for qualitative discussion may be
treated as constants.

Inspection of the normalization integral (\ref{NBS}) will provide
essential guidance to an appropriate choice of the quasipotential propagator.
For that purpose we express the integrand of the normalization integral
explicitly as a function of $\tilde p$ $\hat P$, $k^0$ and $M$.
Since
\beq
p_1= \tilde p +\left(\half M +k^0\right)\hat P \; ,\qquad
 p_2=-\tilde p +\left(\half M  -k^0\right)\hat P\; ,
\eeq
the  nucleon propagators (\ref{PROP}) take the form
\beqarray
&&S(p_1)=-[\Lambda^+_1D^+(-k^0)+\Lambda^-_1D^-(k^0)]\; ,\cr\cr
&&S(p_2)=-[\Lambda^+_2D^+(k^0)+\Lambda^-_2D^-(-k^0)]\; ,
\label{SSG}
\eeqarray
when expressed as functions of $\tilde p$ $\hat P$, $k^0$ and $M$.
Here we use the notation
\beq
D^+(k^0):= {1\over 2\pi i}\,{1\over k^0+(\omega-\half M)-i\epsilon} \; ,\qquad
D^-(k^0):={1\over 2\pi i}\,{1\over k^0+(\omega+\half M)-i\epsilon}.
\eeq
The Dirac spinor matrices $\Lambda_k^\pm$, defined by
\beq
\Lambda_1^\pm:=
{-\gamma^{(1)}\cdot \tilde p \pm\omega \beta^{(1)} +m
\over 2\omega}\;, \qquad
\Lambda_2^\pm:=
{\gamma^{(2)}\cdot \tilde p \pm\omega \beta^{(2)} +m
\over 2\omega}\; ,
\label{PROJ}
\eeq
with $\beta^{(k)}:= -\gamma^{(k)}\cdot \hat P$,
project onto subspaces with
positive and negative scalar products respectively:
\beq
 \Lambda_k^\pm \beta^{(k)} \Lambda_k^\pm =\pm \Lambda_k^\pm
\; ,\qquad
 \Lambda_k^+ \beta^{(k)} \Lambda_k^-
= \Lambda_k^- \beta^{(k)} \Lambda_k^+=0\;  .
\eeq
Note that $(\Lambda_k^+ + \Lambda_k^-)\beta^{(k)}=1$.

In this notation two-nucleon propagator $G_0(p,P)=S(p_1)S(p_2)$
appears in a form which emphasizes the
role of the spinor projectors,
\beqarray
&&G_0(p,P)={1\over 2\pi i}\Biggl\{{\Lambda^+_1\Lambda^+_2\over 2\omega-M}
\left[D^+(-k^0)+D^+(k^0)\right]
 +{\Lambda^-_1\Lambda^-_2\over 2\omega+M}
\left[D^-(-k^0)+D^-(k^0)\right]\cr\cr
&&+{\Lambda^+_1\Lambda^-_2\over M}
\left[D^+(-k^0)-D^-(-k^0)\right]
+{\Lambda^-_1\Lambda^+_2\over M}
\left[D^+(k^0)- D^-(k^0)\right]\Biggr\}\; .
\eeqarray
When the deuteron vertex is $\Gamma_P(\tilde p, k^0)$ is independent of $k^0$,
the integral over $k^0$ in the normalization integral can now be carried
out  explicitly.
{}From
\beq
\int dk^0 S(p_1)\beta^{(1)}G_0(p,P)= {1\over (2\pi)^2}
\left[{\Lambda_1^+\Lambda_2^+\over (2\omega -M)^2}-
{\Lambda_1^-\Lambda_2^-\over (2\omega +M)^2}\right]
\eeq
follows the normalization of the deuteron vertex,
\beq
{1\over 8\pi^2 P^0}\int d^3\tilde p \bar \Gamma_P(\tilde p)
\left[{\Lambda_1^+\Lambda_2^+\over (2\omega -M)^2}-
{\Lambda_1^-\Lambda_2^-\over (2\omega +M)^2}\right]\Gamma_P(\tilde p) =1\; ,
\label{NNG}
\eeq
which by eq. (\ref{JIA}) also gives the value of the total charge correctly.

If we define the unconstrained BSLT propagator, $\hat g_0$,  by
\beq
\hat g_0(\tilde p, P):= {1\over 2\pi i}
{\Lambda_1^+\Lambda_2^+\over 2\omega -M},
\label{QPROP}
\eeq
the normalization condition for the deuteron vertex implies
the normalization of the quasipotential wave function (\ref{WFNBS}),
$\varphi(\tilde p):= \varphi_P(p)_{k^0=0}$.
Since
\beq
{1\over 8\pi^2 P^0}\int d^3\tilde p \bar \Gamma_P(\tilde p)
{\Lambda_1^+\Lambda_2^+\over (2\omega -M)^2}\Gamma_P(\tilde p) \;
= \int d^3\tilde p \bar \varphi(\tilde p)\beta^{(1)}\beta^{(2)}\varphi(\tilde
p)
= \int d^3\tilde p \vert \varphi(\tilde p)\vert^2\; ,
\eeq
we have
\beq
\int d^3\tilde p \int d^3\tilde p'\bar \varphi(\tilde p')
N(\tilde p';\tilde p)\varphi(p)=1\; ,
\eeq
where $N(\hat p',\tilde p)$ is defined as
\beq
N(\tilde p';\tilde p):= \beta^{(1)}\beta^{(2)}\delta^{(3)}(\tilde p' -\tilde p)
-{1\over 4\pi^2}\int d^3\tilde p''\bar \V(\tilde p';\tilde p''){\Lambda^-_1
\Lambda^-_2\over
 (2\omega+M)^2}
\V(\tilde p'';\tilde p)\; .
\label{NORM}
\eeq
This expression shows explicitly how the suppressed negative energy
components contribute to the normalization integral through the last
term on the right hand side.

The  propagator (\ref{QPROP}), when multiplied by $\delta(k^0)$,
has the required  large-time limit,
\beq
\lim_{t\to\infty}g_0(p,P)\exp i(\omega -\half M)t]
=\half \Lambda_1^+\Lambda_2^+ \delta (k^0)\delta(\omega-\half M)\; .
\label{LIM}
\eeq
which is the same as the large-time limit (\ref{LLIM}) of the unconstrained
two-nucleon propagator. The residue of the pole  at $M=2\omega$ is
determined by the requirement (\ref{LIM}).
Considerable freedom is left for the choice the of other features of
this function \cite{Klein}.
The most common choice  \cite{Lomon}:
\beq
g_{PL}(p, P):= {\delta(k^0)\over 2\pi i}
{\Lambda_1^+\Lambda_2^+\over 2\omega-M}\, {2\omega\over \omega+ \half M}
\label{PL}
\eeq
differs from (\ref{QPROP}) by the factor $2\omega/( \omega+ \half M)$ which can
easily be absorbed into the potential and the wave function.
In ref. \cite{Hummel} the following alternative choice is used:
\beqarray
&&g_{HT}(p, P):= {(\gamma^{(1)}\cdot p_1 -m)
(\gamma^{(2)}\cdot p_2 -m)\over
4\pi i[ \fourth P^2+\tilde p^2+m^2-i\epsilon]}\,
{\delta(k^0)\over \half M+\omega}\cr\cr
&&={\delta(k^0)\over 2\pi i}\Biggl[
{\Lambda_1^+\Lambda_2^+\over 2\omega-M}
+{\Lambda_1^-\Lambda_2^-\over 2\omega+M}\,{\omega-\half M\over \omega+\half M}
+{\Lambda_1^+\Lambda_2^- +\Lambda_1^-\Lambda_2^+\over 2\omega+M}\Biggr]
\; .
\label{HT}
\eeqarray
Finally the propagator  recently proposed by Mandelzweig and Wallace
\cite{Wallace} is in our notation
\beq
g_{MW}(p, P):={\delta(k^0)\over 2\pi i}
\Biggl[{\Lambda_1^+\Lambda_2^+\over 2\omega-M}
+{\Lambda_1^-\Lambda_2^-\over 2\omega+M}
+{\Lambda_1^+\Lambda_2^- +\Lambda_1^-\Lambda_2^+\over 2\omega }\Biggr]\; ,
\label{MW}
\eeq
These propagators  include negative-energy  spinor components in the
quasipotential wave functions in different ways. For the purposes of our
discussion the  propagator (\ref{QPROP}) is the most
convenient as it leads to the most explicit interpretation of the
exchange currents  that arise  in the quasipotential reduction.\\
\vspace{0.3cm}

\noindent{\bf 2.3 The BSLT impulse approximation and  exchange currents}
\vspace{0.2cm}

In order to isolate  in eqs. (\ref {JIA}) and (\ref {JIAC}) a term
that represents
the impulse contribution in the quasipotential framework we
use the identities
\beq
G_0(p,P_i)\Gamma_{P_i}(p)\equiv
\delta(k^0)\varphi_{P_i}(\tilde p,0){\sqrt{2P_i^0}} +
\Delta(p,P_i)\Gamma_{P_i}(p)\; ,
\eeq
and
\beq \bar \Gamma_{P_f}(p')S(p_1')\equiv{\sqrt{2P_f^0}} \bar \varphi(\tilde
p',{k'}^0)
{\beta'}^{(2)}+\bar \Gamma_{P_f}(p')\Delta_1(p',P_f),
\eeq
where ${\beta'}^{(k)}:= -{\gamma '}^{(k)}\cdot \hat P_f$.
The propagator complements
$\Delta(p,P)$ and $\Delta_1(p,P)$ are defined as
\beq
\Delta(p,P):= G_0(p,P)-g_0(p,P) = \Delta^+(p,P)+ \Delta^-(p,P)\; ,
\eeq
and
\beq
\Delta_1(p,P):=S(p_1)+\hat g_0(p,P)\beta^{(2)}=
\Delta_1^+(p,P)+\Delta_1^-(p,P)\; .
\label{DDD}
\eeq

Here $\Delta^+$ and $\Delta_1^+$ are projections into positive-energy
components
\beq
\Delta^+(p,P):=\hat g_0[D^+(k^0)+D^+(-k^0)-\delta(k^0)]\; ,
\eeq
and
\beq
\Delta_1^+(p,P):=-\hat g_0{\omega-\half M + k^0\over \omega-\half
 M-k^0-i\epsilon}\beta^{(2)}\, ,
\eeq
while $\Delta^-$ and $\Delta_1^-$ are the remainders which involve
negative-energy components for at least one of the nucleons:
\newpage
\beqarray
&&\Delta^-:={1\over 2\pi i}\Biggl\{{\Lambda^-_1\Lambda^-_2\over 2\omega+M}
\left[D^-(k^0)+D^-(-k^0)\right]
+{\Lambda^+_1\Lambda^-_2\over M}
\left[D^+(-k^0)-D^-(-k^0)\right] \cr\cr
&&+{\Lambda^-_1\Lambda^+_2\over M}
\left[D^+(k^0)-D^-(k^0)\right]\Biggr\}\; , \cr\cr
&&\Delta_1^-(p,P):=-\Lambda_1^+\Lambda_2^- \beta^{(2)} D^+(-k^0)
+\Lambda_1^-D^-(k^0)\;.
\label{DEL}
\eeqarray
The desired isolation of a reduced BSLT impulse term
can now be achieved by means of the identity
\beqarray
&&S(p'_1)K^{(1)}G_0(p,P_i)
\equiv -\hat g_0(p',P_f){\beta'}^{(2)}K^{(1)}g_0(p,P_i)\cr\cr
&&-\hat g_0(p',P_f){\beta'}^{(2)}K^{(1)}\Delta(p,P_i)
+\Delta_1(p',P_f)K^{(1)}G_0(p,P_i)\; ,
\eeqarray
which yields the following separation of the single nucleon
current matrix element (\ref{JIA}):
\beq
\<P_f|J^{IA}_\mu(0)|P_i\>=\<P_f|J^{ia}_\mu(0)|P_i\>+\<P_f|J^{rel}_\mu(0)|P_i\>
+\<P_f|J^{ex}_\mu(0)|P_i\>\; .
\label{SEP}
\eeq
The first term
\beq
{\<P_f|J^{ia}_\mu(0)|P_i\>\over 2{\sqrt{P_f^0P_i^0}}}
:= \int d^3\tilde p \int dk^0
\bar \varphi(\tilde p',{k'}^0){\beta'}^{(2)}K^{(1)}\delta(k^0)\varphi(\tilde
 p,k^0)
+\;(1\leftrightarrow 2)\; ,
\label{IMP}
\eeq
represents the impulse
approximation in the BSLT framework. Here $\tilde p'$ and ${k'}^0$ are
understood to be functions of $\tilde p$ and $k^0$ according to eq.
(\ref{CON}).
The
second term,
\beqarray
&&{\<P_f|J^{rel}_\mu(0)|P_i\>\over 2{\sqrt{P^0_fP^0_i}}}:=\cr\cr
&&\int d^3\tilde p\int dk^0
\Bigl\{\bar \varphi(\tilde
p',{k'}^0){\beta'}^{(2)}K^{(1)}[D^+(k^0)+D^+(-k^0)-\delta(k^0)]\varphi(\tilde
 p,k^0)
\cr\cr
&&+\bar \varphi(\tilde p',{k'}^0)\,{\omega'-\half M +{k'}^0\over \omega' -\half
M-{k'}^0-i \epsilon}{\beta'}^{(2)}\,K^{(1)}[D^+(k^0)+D^+(-k^0)]\varphi(\tilde
 p,k^0)\Bigr\}\cr\cr
&&+\;(1\leftrightarrow 2)\; ,
\eeqarray
represents relativistic effects that vanish for $Q^2=0$.
With instantaneous
quasipotentials this
term is negligible when the $k^0$ dependence of $\tilde p'$ can be neglected.

The last term
\beqarray
&&{\<P_f|J^{ex}_\mu(0)|P_i\>\over 2{\sqrt{P^0_fP^0_i}}}:=\cr\cr
&&\int d^3\tilde p\int dk^0
\Bigl\{\bar \varphi(\tilde p',{k'}^0)
{2\omega'-M\over \omega'-\half
M-{k'}^0-i\epsilon}{\beta'}^{(2)}K_\mu^{(1)}\Delta^-(p,P_i)
\V_{P_i}\varphi(\tilde p, k^0)\cr\cr
&&+{\overline{ \V_{P_f} \varphi}}(\tilde p',{k'}^0)
 \Delta^-_1(p',P_f)K_\mu^{(1)}
[D^+(k^0)+D^+(-k^0)]\varphi(\tilde p,k^0)
\cr\cr
&&+{\overline{ \V_{P_f} \varphi}}(\tilde
 p',{k'}^0)\Delta^-_1(p',P_f)K_\mu^{(1)}\Delta^-(p,P_i)
\V_{P_i}\varphi(\tilde p, k^0)
\Bigr\}\cr\cr
&&+\;(1\leftrightarrow 2)\; ,
\label{EXCH}
\eeqarray
involves
negative-energy spinor components of the wave functions and the large energy
denominators
characteristic of the so-called pair currents \cite{Blunden.a,Blunden.b}.
Here we have used the notation
\beq
\V_{P_i} \varphi(\tilde p,k^0) \equiv \int d\tilde p'' \V_{P_i}(\tilde
p,k^0;\tilde p'',0)\varphi (\tilde p'', 0).
\eeq
This decomposition of the impulse approximation matrix element into
a reduced impulse approximation and exchange current components
depends on the choice (2.53) for the  quasipotential propagator.
The alternate quasipotential choices (2.59) and (2.60) would lead
to different decompositions, and to less obvious interpretations
of the associated exchange current contributions. In the
calculation of the electromagnetic form factors of the deuteron
in  ref. \cite{Hummel}, which was based on the propagator (2.59)
only the associated impulse approximation current matrix element
was taken into account.

The additional
dynamical exchange currents required
by current conservation are  considered in the next section.\\

\newpage
\sect{Current conservation and the minimal exchange current}
\setcounter{equation}{0}
\vspace{3mm}
Gauge covariance of the single-nucleon propagator in the presence of
electromagnetic fields implies the requirement that the
 three-point  current vertex,  $ K_\mu^{(1)}(p_1';  p_1)$, must satisfy the
 the Ward-Takahashi identity
\cite{Ward,Taka}
\beq
(p_1'-p_1)^\mu K_\mu^{(1)}(p_1';  p_1)={1\over2} (1+\tau_3^{(1)})[S^{-1}
(p_1') - S^{-1} (p_1)]\; .
\label{WT}
\eeq
The usual phenomenological form for extended nucleons (\ref{FAC})
violates this requirement. In ref. \cite{Gross} it was pointed
out that this defect can be cured by assuming the  general form
\beqarray
K^{(1)\mu}(p_1';  p_1)&=& \gamma^{(1)\mu}+
\left[\gamma^{(1)\mu} -{Q^\mu (\gamma^{(1)}\cdot Q)\over Q^2}\right]
\left\{\tilde{F}_1^{(1)}(Q^2)-1\right\}\cr\cr
&&+\frac {1}{4m}[\gamma^{(1)\mu},\gamma^{(1)\nu}]Q_\nu \tilde{F}
_2^{(1)}(Q^2)\; .
\label{KGR}
\eeqarray
Since the divergence
of the last two term vanishes identically  the  the modified current vertex
(\ref{KGR}) manifestly satisfies the Ward-Takahashi identity.\\

When the  Ward-Takahashi identity (\ref{WT}) is
applied to the relativistic impulse approximation (\ref{JIA}) for the
current   we find that
\beqarray
&&Q^\mu \<P_f|J^{IA}_\mu(0)|P_i\>
= \int d^4p'\int d^4p  \delta(p'-p-\half Q)\cr\cr
&&\times \bar\Gamma_{P_f}(p')\frac{1+\tau_3^{(1)}}{2}
\left[G_0(p,P_i)-G_0(p',P_f)\right]\Gamma_{P_i}(p)
+(1 \leftrightarrow 2) \; .
\label{IAC}
\eeqarray
This result shows that the impulse approximation does, in general, not satisfy
current conservation.  Therefore the presence of
an exchange current matrix $\<P_f|J^{EX}_\mu(0)|P_i\>$ is required such that
the full current is conserved,
\beq
Q^\mu \<P_f|J^{IA}_\mu(0)|P_i\>+Q^\mu\<P_f|J^{EX}_\mu(0)|P_i\>=0\; .
\eeq

When  $G_0$ is replaced by $g_0+\Delta$ in the matrix
element (\ref{IAC})
the BSLT equation (\ref{VV}) yields the result
\newpage
\beqarray
&& Q^\mu\<P_f|J_\mu^{IA}(0)|P_i\>=\cr\cr
&&\int d^4 p\int d^4p'
\bar\Phi_{P_f}(p')\,\Biggl\{ \bar\V_{P_f}(p';p+\half Q)
\frac{1+\tau_3^{(1)}}{2}
- \frac {1+\tau_3^{(1)}} {2}
 \V_{P_i}(p'-\half Q;p) \cr\cr
&&+ \int d^4p'' \bar \V_{P_f}(p',;p''+\half Q)\frac {1+\tau_3^{(1)}} {2}
\left[\Delta(P_i,p'')- \Delta(P_f,p''+\half Q) \right]\V_{P_i}(p'',p)
\Biggr\}\,\Phi_{P_i}(p)\cr\cr
&& + (1 \leftrightarrow 2)\; ,
\label{IAI}
\eeqarray
where
\beq
\Phi_{P}(p):= g_0(p,P)\Gamma_P(p)= \delta(k^0)\varphi(\tilde p,k^0){\sqrt{
2P^0}}\; .
\eeq

A ``minimal'' exchange   current vertex $\hat K_0^{EX}$ that suffices to
restore
current conservation  has the form
\beqarray
&&\<P_f|J^{EX}_\mu(0)|P_i\> =
- \int d^4p' \int d^4p\frac{q}{q_\mu\cdot Q}\cr\cr
&&\times\bar \Phi_{P_f}(p'),\Bigg\{
\Bigl[\bar{\cal V}_{P_f} ( p'\; ;  p+\half Q )\frac {1+\tau_3^{(1)}} {2}
-\frac {1+\tau_3^{(1)}} {2}{\cal V}_{P_i} (p'-\half Q ;p)
\Bigr]
\cr\cr
&&+ \int d^4p'' \bar \V_{P_f}(p';p''+\half Q)\frac {1+\tau_3^{(1)}} {2}
\Bigl[\Delta(P_i,p'')- \Delta(P_f,p''+\half Q ')\Bigr]
\V_{P_i}(p'';p)\Biggr\}\,\Phi_{P_i}(p)\cr\cr
&& +(1\leftrightarrow 2)
\; .
\label{KEXI}
\eeqarray
where $q:=p'-p-\half Q$. Only the longitudinal part of this exchange current
is determined by the requirement of current conservation. The transverse part
of the exchange current is however consistent with meson-exchange
models \cite{Gross}. Since the right-hand side
of eq.~(\ref{IAC})  is independent of the nucleon form factors
the ``minimal'' exchange currents are obviously independent of the nucleon
structure \cite{Gross}.
One can, of course, add any purely transverse two-body
current to the ``minimal'' exchange current, $\hat K_0^{EX}$.
Such  ``model dependent'' additions should be
motivated by models of the underlying structure. Consideration
of meson-exchange models suggest \cite{Gross} the modification of the
transverse
 part
by a factor representing meson structure,
\beq
J_\mu^{EX} \to J_\mu^{EX}+ [J_\mu^{EX}-\frac{Q_\mu}{Q^2}(Q \cdot
 J^{EX})][F_M(Q)-1] \; ,
\label{KEXT}
\eeq
where $F_M(Q^2)$ is an arbitrary mesonic form factor.

The interpretation of the two terms
in the ``minimal'' exchange current
(\ref{KEXI}) is straightforward. The first term, which  is linear in the
quasipotential, represents the
exchange current  that is associated with the isospin
dependence and the non-locality of the quasipotential ${\cal V}$. It would
vanish in the case of an isospin independent quasipotential that
depends only on the invariant momentum transfer. Such a term
term  is  required, for the same reasons  in the framework
of the Bethe-Salpeter formalism \cite{Gross}.  The quadratic term
arises specifically from  the quasipotential reduction. It vanishes in the
static
 limit.

In practical calculations it is necessary to combine the minimal
exchange current  (\ref{KEXI}) with the effective two-body current
 (\ref{EXCH}) which arose from the quasipotential reduction.
The sum of the exchange current matrix elements  (\ref{KEXI}) and
 (\ref{EXCH}) gives the total exchange-current matrix element
that is appropriate to  the BSLT framework.\\

\sect{Phenomenological exchange currents}
\setcounter{equation}{0}
\vspace{3mm}

Exchange current effects are known to be significant in a number
of electronuclear reactions. The relative magnitude
of the exchange currents typically increases
with momentum transfer \cite{Riska.a}. It is therefore important to describe
exchange currents   by methods
that respect both Lorentz-covariance and current conservation. In most
applications  exchange current operators have however been derived
on the basis of  nonrelativistic approximations to Fock-space representations
of quantum field theory. Such approximations  cannot be expected
to be reliable for large values of the momentum transfer.\\

The Bethe-Salpeter formalism provides one approach
to overcome these limitations. The problem then is  to establish contact with
the quantum mechanical description of bound-state structure. For this
purpose  we have  exploited covariant quasipotential reductions to derive
 effective exchange currents (\ref{EXCH}) and (\ref{KEXI}).
Here we will  compare our relativistic
expressions with  the usual nonrelativistic phenomenological results.
For simplicity we assume
\newpage
\noindent that ${\cal V}$ depends solely on the invariant 4-momentum
transfer  $ q^2:=( p'- p)^2$,
\beq
{\cal V}_P(p';\;p)=\V(p'-p)=v^+( q^2)+v^-( q^2)\vec \tau^1 \cdot \vec
\tau^2\; ,
\label{POT}
\eeq
where $q:=p'-p$. The isospin dependence has been made
explicit in order to illuminate the relation between isospin
dependence and the exchange currents.\\

The  second term in eq. (\ref{IAI}) can be  written in the form
\beq
\int d^3\tilde p\int dk^0 \bar \Gamma_{P_f}(\tilde p',0)
\left[\Delta(P_i,\tilde p, k^0)-\Delta(P_f,\tilde p', {k'}^0)\right]
\Gamma_{P_i}(\tilde p,k^0)\; ,
\eeq
where $\tilde p '$ and ${k'}^0$ are related to $\tilde p$ and $k^0$ by the
impulse constraint (\ref{CON}). In this form it is evident that this term
vanishes in the static limit and remains small for moderate values of $Q^2$.
We therefore neglect the contribution of this term to the exchange current
(\ref{KEXI}) for the following discussion.

With the assumption (\ref{POT})  for the potential the exchange current
defined in eq. (\ref{KEXI}) then reduces to the form
\beq
{\<P_f|J^{EX}_\mu(0)|P_i\>\over 2 {\sqrt{P^0_fP^0_i}}}
=i\int d^3\tilde p' \int d^3 \tilde p
\bar \varphi(\tilde p',0)(\tau^{(1)}
\times \tau^{(2)})_3{\tilde q_\mu \over\tilde q\cdot Q}
v^-(\tilde q^2)\varphi(\tilde p,0)
\label{KEXP}
\eeq
where $\tilde q:= \tilde p'- \tilde p$.
This exchange current  is formally identical to that employed in
the literature for the  currents associated
with exchange of mesonic systems with isospin 1
\cite{Riska.b,Tsushima,Kirchbach}. \\

The other component of the required exchange current  is the pair current
 $J_\mu^{ex}$, defined in eq.(\ref{EXCH}), which arises as a consequence
 of the
quasipotential reduction.
In the static limit the
first and second  term of the  exchange current
(\ref{EXCH})
involve
only the components $\Lambda_1^+K^{(1)}\Lambda_1^-$ and
$\Lambda_1^-K^{(1)}\Lambda_1^+$ respectively.
Thus only matrix elements of $K_\mu^{(1)}$ that connect the large and small
components contribute to these terms. In the last term there
are also contributions diagonal in the small components which contribute to
the normalization (\ref{NORM}) of the BSLT wave function. The result is
\newpage
\beqarray
&&{\<P_f|J^{ex}_\mu(0)|P_i\>\over 2{\sqrt{P^0_fP^0_i}}}\approx
{1\over 4\pi i m} \int d^3\tilde p
\Biggl\{\bar \varphi(\tilde p')
K_\mu^{(1)}\Lambda_1^- \V_{P_i}\varphi(\tilde p)
+{\overline{ \V_{P_f} \varphi}}(\tilde p')
\Lambda_1^-K_\mu^{(1)} \varphi(\tilde p)
\cr\cr
&&-{1\over 4\pi^2}{\overline{ \V_{P_f} \varphi}}(\tilde p')
\left[{\Lambda_1^+K_\mu^{(1)}\Lambda_1^-\Lambda_2^- +\Lambda_1^-K_\mu^{(1)}
\Lambda_1^+\Lambda_2^-\over 2(2m)^2}-{\Lambda_1^-K_\mu^{(1)}
\Lambda_1^-\Lambda_2^- \over (4m)^2}\right]
\V_{P_i}\varphi(\tilde p)
\Biggr\}\cr\cr
&&+\; (1 \leftrightarrow 2) \; .
\eeqarray
The first two terms on the r.h.s. have the same form as the pair
currents that arise in the conventional quantum mechanical
framework. The last term represents the electromagnetic current
of the suppressed negative energy components.
This derivation
clearly exhibits the approximations that underlie the conventional
form of the pair current operators.  \\


\sect{Discussion}
\setcounter{equation}{0}
\vspace{3mm}

The derivation above of the effective exchange currents
to be used in the quasipotential formalism  is
general enough to be applied to the axial current of the
two-nucleon system. In the
case of the axial current  the impulse approximation and
effective exchange currents  are given by the general
expressions (\ref{IMP}) and (\ref{EXCH}), once the kernel $K_\mu^{(1)}$ is
taken
to be the axial current kernel for a single nucleon. These expressions will
then permit a more systematic  approach to the
derivation of the effective axial pair currents, which are
known to be important in first forbidden nuclear $\beta$-transitions
\cite{Kirchbach,Towner} and pion production reactions \cite{Lee}.
Since axial currents are not
conserved there is no analog to the dynamical
exchange current (\ref{KEXI}).\\

Some  remarks are in order concerning comparisons of BS-BSLT phenomenology
with Hamiltonian quantum mechanics. The relations are tenuous at best.
The formal relations to equal-time Fock-space
representations of the quantum field theory exist only perturbatively.
They involve the equal-time constraints, discussed in the early history of the
subject by  Levy, Klein and Macke \cite{Klein}, rather than the BSLT
constraints which have prevailed more recently. Quantum field theory describes
system of infinitely many degrees of freedom and physical nucleons are
necessarily composites. The reduction to an effective two-nucleon dynamics
is thus qualitatively similar to the treatment of atoms and
molecules as inert particles.  This analogy also illuminates the role of the
static limit. Our fully relativistic result has been cast into a form in which
the  static limit allows  comparison with nonrelativistic treatments.
 On the other hand it is
obvious from the definition of the BS wave functions that
there is no simple functional relation to the
square integrable functions representing states in
relativistic Hamiltonian dynamics \cite{CCKP,keister}. In that formulation
of dynamical models  the number of degrees of freedom is restricted at the
outset and  an
explicit unitary representation  of the Poincar\' e group in the two-nucleon
Hilbert
space provides both the dynamics and the Lorentz transformations.
Current conservation and Lorentz covariance imply dynamical constraints on the
current operators.\\

The main results of this paper are based on the
the exact Mandelstam relation (\ref{JGAM}) together with
vertex identity (\ref{EQG}). On that basis we derived an explicit expression
for the effective exchange current (\ref{EXCH})  without approximations.
No assumptions have been made in the derivations of the formulae in this
paper on the explicit form of the quaspotential ${\cal V}$. Thus
any unconstrained polyzouquasipotential fitted to two-nucleon data may be used.
Boson exchange models have the virtue of relating interaction currents to the
exchange of bosons and lead to a simple dynamical interpretation
of the different components of the effective exchange current.
Such models serve to motivate the otherwise arbitrary
transverse part of the dynamical exchange current (\ref{KEXT}).\\

\centerline{\bf Acknowledgement}
\doeack
\newpage
\appendix
\sectA{Appendix}

The four-vector $\tilde p$ is a function of three independent variables
\beq
 k_a :=u_a(P)\cdot p\; ,
\eeq
where $u_a(P)$ are three orthonormal vectors which are orthogonal to
$\hat P$:
\beq
u_a(P)\cdot u_b(P)= \delta_{a,b}\; , \qquad
u_a(P)\cdot \hat P =0.
\eeq

The projections of the relative four-momenta  $p$ and $p'$ into the
hyperplane spanned by the initial and final four-momenta $P_i$ and
$P_f$ can be  expressed as linear combinations of orthonormal
four vectors. Different choices for this basis may be convenient for different
purposes.
For instance
\beqarray
&&p= k^0 {P_i\over M_i} + k_Q {\tilde Q_i
\over {\sqrt {\tilde Q_i^2}}} + p_\perp\cr\cr
&&p'={k'}^0 {P_f\over M_f} + k'_Q {\tilde Q_f \over {\sqrt {\tilde Q_f^2}}} +
p'_\perp
\eeqarray
where $\tilde Q_i$ and $\tilde Q_f$  are projections of $Q$ perpendicular to
$P_i$ and
$P_f$ respectively,
\beq
\tilde Q_i:=Q- {P_i\cdot Q \over P_i^2}\, P_i\; , \qquad
\tilde Q_f:=Q- {P_f\cdot Q \over P_f^2}\, P_f\; ,
\eeq
and the subscript $_\perp$ indicates the projections orthogonal to the
hyperplane
spanned by $P_i$ and $P_f$. A basis symmetric in the initial and final states
is
specified by the orthogonal unit vectors $\hat Q:= Q/{\sqrt{Q^2}}$ and $\hat
P_B:=P_B/M_B$,
\beq
P_B:= \half [P_i+P_f]- {(P_i+P_f)\cdot Q\over2 Q^2}\, Q\; ,
\eeq
\beq
M_B^2= -P_B^2= \half [M_f^2+M_i^2]+\fourth Q^2 +{(M_f^2-M_i^2)^2\over 4 Q^2}\;
{}.
\eeq
In this basis the momentum constraint of the impulse approximation,
$p'-p=\half Q$, takes the simple form,
\beq
p'_\perp =p_\perp\; , \qquad  {k'}_B^0=k_B^0\; , \qquad
k'_{BQ}-k_{BQ}= \half {\sqrt{Q^2}}\;,
\eeq
where the four Lorentz scalars ${k_B'}^0$, $k_B^0$ , $k_{BQ}$ and $k'_{BQ}$ are
defined
by
\beq
p= k_B^0 \hat P_B +k_{BQ}\hat Q +p_\perp\; , \qquad
p'= {k'}_B^0 \hat P_B+{k'}_{BQ}\hat Q +p_\perp\; .
\eeq

The Jacobian of the variable transformation $k^0,k_Q \to k^0_B,k_{BQ}$ is
unity. We have
\beq
d^4p = d^2p_\perp dk_Q dk^0 = d^2p_\perp dk_{BQ} dk_B^0\; ,
\eeq
and
\beq
\delta^{(4)}(p'-p-\half Q)=\delta^{(2)}(p'_\perp-p_\perp)
\delta(k'_{BQ}-k_{BQ}-{\sqrt{\eta}})\delta({k'}^0_B-k^0_B)\; .
\eeq


\end{document}